# THE ETHICAL IMPLICATIONS OF DIGITAL CONTACT TRACING FOR LGBTQIA+ COMMUNITIES

Izak van Zyl, Cape Peninsula University of Technology, vanzyliz@cput.ac.za

Nyx McLean, Cape Peninsula University of Technology, mcleann@cput.ac.za

**Abstract:** The onset of COVID-19 has led to the introduction of far-reaching digital interventions in the interest of public health. Among these, digital contact tracing has been proposed as a viable means of targeted control in countries across the globe, including on the African continent. This, in turn, creates significant ethical challenges for vulnerable communities, including LGBTQIA+ persons. In this research paper, we explore some of the ethical implications of digital contact tracing for the LGBTQIA+ community. We refer specifically to the digital infringement of freedoms, and ground our discussion in the discourse of data colonisation and Big Tech. We propose a critical intersectional feminism towards developing inclusive technology that is decentralised and user controlled. This approach is informed by a feminist ethics of care that emphasises multiple lived experiences.

## 1. INTRODUCTION

In this research paper, we respond to Track 7 of the IFIP WG 9.4 2021 conference, namely, *Feminist and Queer Approaches to Information Systems (IS) in Developing Countries.* Taking a continental (African) perspective, we argue that the technological containment strategies around the COVID-19 pandemic create serious implications for vulnerable populations in the LGBTQIA+ (lesbian, gay, bisexual, transgender, queer/questioning, intersex, asexual/ally, and other) community. Particularly, we argue that the widespread introduction of contact tracing mechanisms by governments and affiliated organisations infringes on the privacy, freedom, and rights of these and other groups. This occurs through obtaining – and potentially exploiting – sensitive information that could otherwise be used to denigrate, punish, and physically harm or kill vulnerable people. We frame this discussion within the broader discourse of data colonialism, where Big Tech assumes its newfound role as public health policymakers (Sharon, 2020).

In what follows, we review the seminal literature in the domain of digital ethics and Big Tech in the context of the current pandemic. Specifically, we examine digital contract tracing and its intrusive possibilities for members of the LGBTQIA+ community across Africa. We refer here to Africa not only in a geographic sense, but primarily in the cultural historical sense of it being a casualty of deliberate underdevelopment through colonisation (Rodney, 2018). We discuss the historical persecution of vulnerable communities on the continent and argue against the current and potential digital infringement on LGBTQIA+ freedoms. Furthermore, we critique the role and influence of big technology companies in this context. We conclude the paper by highlighting theoretical and practical approaches through which to promote collaborative and empathetic technological development with LGBTQIA+ communities.

## 2. DIGITAL CONTACT TRACING AND PRIVACY

With the global onset of COVID-19, governments have introduced far-reaching measures to contain the virus. These have included contact tracing as a familiar means to fight infectious disease outbreaks, as historically in cases like Ebola and SARS. Contact tracing, in principle, involves highly targeted control, whose purpose is to prevent further infection. Often coupled with treatment and isolation, contact tracing is a severe form of targeted control and is typically effective in cases with low rates of infection (Eames & Keeling, 2003). According to Dar and colleagues (2020, p.2), contact tracing conventionally occurs in three steps or phases:

1) *Identification*: identify those persons an infected patient came into personal contact with.

2) *Listing*: maintain a record of the contacts of the infected patient(s) and inform those persons.

3) *Follow-up:* further examination, treatment, and isolation of contacts, and especially those who test positive.

Contact tracing is not always a good control strategy. It may be less successful in high-risk and heterogeneous groups, with many potential transmission routes and a high incidence of infection (Eames & Keeling, 2003). The authors go on to indicate that, "[i]f the tracing process is significantly slower than the infection process then, no matter how large a proportion of contacts is eventually traced, it will be impossible to keep pace with the epidemic; there is a trade-off between tracing speed and tracing efficiency" (ibid., p.2570).

In their analysis on the effectiveness of contact tracing strategies for COVID-19, Kretzschmar and colleagues (2020) recognise that typically, (manual) tracing efficiency is reduced through testing delays that may occur in especially low-resource or dense population settings. However, they find that such delays could be minimised through app-based (digital) technology, thus enhancing contact tracing effectiveness and coverage. To this end, the World Health Organization (WHO) describes distinct categories of digital tools: outbreak response tools, proximity trackers, and symptom tracers (WHO, 2020). Notwithstanding their benefits, the inherent challenges with such digital tools include high infrastructural costs, low overall adherence (participation), and importantly, privacy and data ownership concerns (Kretzschmar et al., 2020; WHO, 2020).

Anglemyer and colleagues (2020) conducted a rapid review on the effectiveness of digital contact tracing versus manual tracing, not limited to COVID-19. In their assessment, they found only low-certainty evidence of success in terms of reduction and prevention of cases. The researchers note the additional risks of privacy breaches, as well as equity concerns for at-risk populations with limited access to resources. Dar et al. (2020) note the benefits of digital contact tracing in light of COVID-19, especially in speeding up the process of identification. However, the researchers raise significant privacy considerations, including "securing the identity of an infected individual from others, stopping the spread of misinformation, … and withholding the countries from establishing a surveillance state" (ibid., p.3). In terms of the latter factor, the authors point to legislation in countries like Israel and South Korea that allows for government tracking of contacts - which involves recording details about occupation, age, travel routes, and importantly, gender.

In Africa, countries like Egypt, Rwanda, South Africa, Democratic Republic of the Congo (DRC), Mozambique and Tanzania have deployed GPS-enabled digital contact tracing, leveraging, inter alia, cell phone tower data, satellite technology, and mobile apps, to fight COVID-19 (Arakpogun et al., 2020). Worryingly, the Ugandan government has sanctioned legal instruments that allow medical officers or health inspectors to enter any premises to search for potential COVID-19 cases, or to inquire about the whereabouts of suspected positive patients (CIPESA, 2021). In the DRC and Tanzania, digital contact tracing may involve unsolicited home visits by community health workers (Nachega et al., 2020).

Ultimately, digital contact tracing poses significant risks to privacy, both in Africa and globally. Governments and health organisations are not necessarily transparent in how data is processed, shared, or protected (see Angelopoulos et al., 2017). Proximity tracking applications especially may disclose location history, case and contact status, and potentially other personal data (WHO, 2020). Taken together, otherwise well-intentioned digital tools may be exploited in the establishment of widespread surveillance states, for example, in Israel (Amit et al., 2020), South Korea (Park, Choi, & Ko, 2020), and non-democracies like China, Jordan, and Vietnam (Greitens, 2020).

## 3. PERSECUTION OF LGBTQIA+ GROUPS IN AFRICA

In Africa, the state of LGBTQIA+ rights is dismal. At the time of writing, seven out of 54 countries protect against discrimination based on gender identity and sexual identity; three have legally recognised gender identity; LGBT rights are legal in 21 countries, and South Africa is the only country to recognise adoption and civil unions. Same-sex marriage is banned in nine out of 54 countries. In Uganda, Tanzania, and Sierra Leone, homosexuality is punishable with life-term imprisonment; while in Sudan, Somalia, Somaliland, Mauritania, and northern Nigeria, homosexuality is punishable by death.

South Africa is internationally recognised as a champion of LGBTQIA+ rights. Globally, it was the first country to constitutionally protect people irrespective of sexual orientation, and later the first country in Africa to recognise same-sex unions through the passing of the Civil Union Act. Despite this, many LGBTQIA+ South Africans continue to experience homophobic and transphobic violence (McLean & Mugo, 2015, p.17). This may seem surprising because constitutional protections and the increased visibility of LGBTQIA+ people lead to the belief that homophobia and transphobia are no longer a cause for concern in South Africa (Teal & Conover-Williams, 2016, p.18).

Homophobic and transphobic harassment and violence manifest in several ways. This can range from verbal abuse to physical assault, sexual violence, and even murder. LGBTQIA+ people may be further marginalised by poverty and racism (McLean, 2018, 2020; Thoreson, 2008). Reid (2020) provides an example of discrimination and persecution that LGBTQIA+ people face in Africa, including Kenya, Egypt, Ghana, and South Africa, and how these discriminations are amplified by socio-economic/poverty – whereby LGBTQIA+ individuals do not have financial capital to achieve independence from abusive families, and the like. The significance of this lies in the implications of leaking personal information about LGBTQIA+ people who live in violent communities or with homophobic family members, and who cannot afford to be evicted during a pandemic with job and food security being at an all-time low.

Further, it is important to note the frequent occurrence in some African countries where religious leaders hold significant sway, and often attribute drought, poverty, a pandemic like COVID-19, and any other hardship a community endures as "divine punishment" for harbouring homosexuals (Reid, 2020). This makes LGBTQIA+ people targets of harassment and violence in their communities. Violence against LGBTQIA+ people, and in particular transgender people, has increased worldwide (Human Rights Campaign, 2020; McLean, 2020). The spike in hate crime-related murders is a result of a "culture of violence" that emerges from transphobia and intersecting discrimination on the basis of racism, sexism, and homophobia (Human Rights Campaign, 2020).

The internet and digital technologies are often understood to be safer spaces for LGBTQIA+ people, such as many of those who live on the African continent, where offline spaces may be violent and dangerous to these groups (McLean, 2020; McLean & Mugo, 2015). Digital spaces, as McLean and Mugo (2015) and McLean (2020) have shown, can be negotiated, navigated, and managed – such as through closed groups and secure/encrypted communication. Some members of these online safe spaces may form connections and migrate to text-based messaging applications such as WhatsApp, or into offline spaces. However, sharing of contact data and/or sharing physical space details may not be secure, given privacy controversies with some social media platforms and their relationships with state governments.

As McLean and Mugo (2015, p.99) have written, the internet makes it possible for LGBTQIA+ people to connect with each other and to form safe digital spaces. However, it also provides people, communities, and governments with "tools to monitor" LGBTQIA+ people, especially in various African countries with anti-homosexuality laws (ibid.). One recent example from Egypt is that of the imprisonment of Andrew Medhat for "public debauchery" after police had located him through the gay dating app, Grindr (Nigro, 2019). He had not yet met with the man he was intending to meet, but officials argued that his use of the app was sufficient evidence.

As Nigro (2019) writes, LGBTQIA+ people as a group, "long criminalized and pathologized, is often an afterthought when it comes to user privacy and regulations - which has resulted in a precarious digital landscape". The reason for this is that the context in which technology is built is a neoliberal one which does not consider how technologies may exacerbate oppressions such as those founded in race, gender, class, and sexuality (Noble, 2018, p.1). Dating apps, for instance, often make use of location data to link users to each other but while useful in finding a match, it can also put users at risk of harassment and violence from those making use of the apps to target LGBTQIA+ people. While this is an example from a dating app, the risk of exposing users' location data is applicable to the use of tracing and quarantine apps during COVID-19. It is important to protect the data and privacy of users, especially for LGBTQIA+ people who are at risk of discrimination and persecution if their identities are made known to homophobic, transphobic, and queerphobic communities and state governments (Stonewall in Fox, 2019).

## 4.   COVID-19 AND 'DIGITAL INFRINGEMENT' ON LGBTQIA+ FREEDOMS

During the first few months of COVID-19, the Association for Progressive Communication (APC) Women's Rights Programme (WRP) (2020) wrote of the adoption of technology and noted the expansion of surveillance and state power. Specifically, they wrote of:

"...reports from many countries, including those in the global South such as Kenya, Uganda and India, that the government is requiring or asking people to install mobile phone apps that use location data for contact tracing. While some measures of surveillance are sophisticated and reliant on tech, others are about marking the bodies of COVID-19 patients with stamps. Routinely, privacy is being compromised and violated, and we are all frightened enough to let this happen".

Those in the race to adopt and "embrace digital contact tracing" - while a timely response to the pandemic - did not consider "putting laws and policies in place to address the stigma surrounding the epidemic, and to protect the rights of those most marginalized, risks undermining the goal of epidemic control" (Chair, 2020).

As online activity of marginalised groups comes to the attention of state officials, digital security and the protection of personal information becomes a matter of urgent attention. We draw on the work of Trias (2020) and Ranjit (2020) who studied tracing applications in India. Trias (2020) writes of how the app Aarogya Setu, using Bluetooth technology, "can trace the relationships between its users and calculate the infection potentials that each one has. When a new positive case is detected, Aarogya Setu immediately alerts the authorities of all persons who were in contact with that individual. In other words, it grants full authority over the handling of personal data to the state". While, to some, this may seem like a concern regarding privacy and access to personal data, for LGBTQIA+ people, this becomes an additional concern around personal safety.

Privacy needs to be understood beyond the notion of sharing information, to also include the various risks to diverse groups. For instance, as Chair (2020) writes, privacy as seen from a feminist lens focuses on the use of our data as a tool of surveillance, one which controls women and LGBTQIA+ people's "bodies, speech, and activism". Women and LGBTQIA+ people "experience privacy differently from men because of the social perceptions of how men and women should behave, determining the extent of privacy in patriarchal societies. The discussion on gender in contact tracing, for example, has left a lot to be desired as the extra risks that women and marginalised communities face are not sufficiently taken into account" (ibid.).

The use of tracing apps is not only an issue of information privacy but also an issue of "autonomy, dignity, bodily integrity, and equality" (Ranjit, 2020). The data being collected on people's movements may come to be used by governments and corporations "as a means of governance and regulation of what our physical bodies can do" (ibid.). The author provides two examples of tools used in India, that of Quarantine Watch which requires citizens to send hourly selfies to officials to prove that they were in their homes; and the COVA app, which uses geo-fencing and location data to create "a virtual border" (Ranjit, 2020).

The risk of such movement monitoring technologies to LGBTQIA+ people is clear: through confining the body to physical spaces with location data and sharing the data publicly, this group is made even more vulnerable. In India, this resulted in those quarantined being further policed by their neighbours who had access to their personal information through the lists of data that officials had made available (Ranjit, 2020). This is comparable to when Ugandan newspaper Rolling Stone (now defunct) published the names of LGBTQIA+ people in 2010 under the headline "100 Pictures of Uganda's Top Homos Leak. Hang Them". The BBC (2010) reported on this in an article titled "Attacks reported on Ugandans newspaper 'outed' as gay". The murder of Ugandan LGBTQIA+ activist David Kato a few months later was attributed to the publication of the names of LGBTQIA+ people (Gettleman, 2011).

The consequences of the public sharing of personal data will not be the same for everyone, and those who are the most vulnerable in society, such as transgender people, may "face greater stigma and discrimination" (Ranjit, 2020). The author provides an explicit example of how in Hyderabad, posters were put up stating that interactions with transgender people would result in contracting the coronavirus. This illustrates the clear danger to vulnerable and marginalised people in the leaking of personal data.

Data recording in the name of a crisis is also problematic in the long term, post COVID-19, as these actions are not reversed once the crisis has passed. Zuboff (2019) provides the example of digital surveillance technologies such as those used during and after the 9/11 moment in the United States, which were justified by the 'war on terrorism'. However, once the moment had passed, those technologies remained. As Ranjit (2020) writes, "the data now collected by various institutions can be used to re-exert control over our bodies, through ways often unknown, making it hard to question this". Furthermore, they write that data and privacy legislation has not kept up with the development of technologies such as those introduced in response to COVID-19. The emphasis on data as a resource, with a focus on reducing the risk of infection, fails to bear in mind how giving up personal data may later result in implications for bodily integrity and autonomy.

While, as Nigro (2019) highlights the persecution of gay people using dating apps, the crimes committed against LGBTQIA+ were "unintentional on the parts of the apps themselves". This brings to attention the need for developers of technology to consider the multiple and intersecting forms of oppression that people may face, and that they should build technology with the "lives and interests of marginalised communities" in mind (Mohanty, 2005, p.511). By building technology in this manner, developers can consider power and its distribution in a way that accounts for the often "unseen, undertheorized, and left out" (ibid.). However, this would require a move away from current modes of Big Tech development, which are primarily driven by for-profit motivations.

## 5.     DATA COLONISATION, PUBLIC HEALTH, AND BIG TECH

While we recognise that many contact tracing apps are not developed by Big Tech, we limit our discussion to the most dominant and valuable companies due their ubiquity. These include the traditional 'Big Five', namely Amazon, Apple, Facebook, Google, and Microsoft. Indeed, Big Tech such as Apple and Google were some of the first to respond to the COVID-19 crisis offering up a digital contact tracing API (Sharon, 2020). However, these companies are driven by a for-profit model and this brings into question the 'social good' of their offering of a digital solution during the most recent crisis. Digital technologies, especially those developed by Big Tech being "rooted in political economy dynamics", may still come to cause harm even if they are "well-intentioned" (Magalhães & Couldry, 2021, p.344). It is important to note here that there is diversity in the approach to digital contact tracing apps. We are therefore mindful of not making sweeping claims about associated privacy and data protection issues. However, the overall point remains that big technology companies are well-positioned to influence individual freedoms through their market dominance.

Considering that many digital solutions and technologies, as well as their data extraction practices, are designed and developed in the Global North and then "deployed in poorer nations" such as in the Global South, colonialism is the most suitable framework to make sense of these data practices (Magalhães & Couldry, 2021, p.345). Like empires who imposed their notion of civilisation on the countries they colonised, Big Tech has swept in with digital solutions for addressing the current public health crisis of COVID-19 relying on desperation for the uptake of technology. Big Tech's response to the current crisis is also familiar: "[a]s forceful promoters of *technological solutionism,* prioritizing technological answers to a broad range of social, economic, political and environmental questions facing contemporary society, they marginalize less intensively technological but possibly more appropriate responses" (Barendregt et al., 2021).

Undoubtedly, the speed at which COVID-19 moved required a rapid response, and digital contact tracing "provide[d] speed, scale and accuracy" (Sharon, 2020). However, in times of crisis, some civil liberties may be suspended by governments, and "the sharing of sensitive data like one's health status and location can contribute to containing the spread of the virus" (ibid.). Digital contact tracing introduces different risks to the recording of data as opposed to recording data on paper, for instance. When the technology is built at speed and rushed to market without the rigorous analysis and testing that may usually be afforded to software, there is a greater risk of vulnerabilities in the software.

In not testing rigorously, room may be left for hacking devices and tracking software. This software, once installed, may be able to open the user's phone up to commands which may then permit the movement of data, "including all passwords, contacts, reminders, text and voice calls. In addition, the operator could turn on the phone's camera and microphone, use its GPS to track the target" (Desai, 2020). The sharing of location data "can be used to show who a person associates with and to infer what they were doing together at a given time" (Sharon, 2020). One must simply look to the previous example of the publishing of identities and personal details such as residential addresses of LGBTQIA+ people in Uganda to imagine the risk that such vulnerabilities in software could expose the LGBTQIA+ community to. Unfortunately, the debates around privacy and digital contact tracing "have tended to be framed in terms of a trade-off between individual privacy rights and public health" (ibid.). This is problematic because individuals now face a 'privacy paradox': a value dilemma through which they are pressured to sacrifice aspects of their privacy in the (supposed) interest of their health (Rowe, 2020).

Imposing digital solutions on people in a crisis without consideration for their lived experiences is akin to the imposition of values and culture of colonising nations who sought to exploit and extract for profit without concern for the 'colonised'. An alternative to the imposition of digital solutions would be to adopt a participatory approach, underpinned by an intersubjective ethics, which sees that digital solutions to public health crises are community-led (see Van Zyl & Sabiescu, 2020). We argue that this approach is best informed by a critical intersectional feminist perspective, guided by a feminist ethics of care, because it would be inclusive of multiple lived experiences in the design, while considering the risks that the technology may present to vulnerable groups such as the LGBTQIA+ community.

# 6. CONCLUSIONS AND RECOMMENDATIONS

In the following section, we contribute both theoretical and practical recommendations to mitigate the risks of digital contact tracing to the LGBTQIA+ community. Namely, we advocate for a critical intersectional feminist approach which accounts for the lived experiences of the most vulnerable, while critically considering the concentration of power over access to personal data.

## 6.1. A critical intersectional feminist approach to technology

Critical intersectional feminism recognises that systemic oppressions like sexism, racism, classism, ableism, colonialism, and cissexism intersect. Intersectionality shows how discriminations exacerbate each other and that they cannot be separated out (Crenshaw, 1989). A critical intersectional feminist approach to the development of digital solutions will consider the interests and lives of the most marginalised: those who are "unseen, undertheorized, and left out" (Mohanty, 2005, p.511). Such an approach would allow for an analysis of power and exploitation that is not evident in the production of solutions by Big Tech. This is because, at the heart of feminism, is the work of disrupting and destabilizing oppressive power structures and dynamics to create a more equal, inclusive, and socially just world (Cooky, Linabary, & Corple, 2018; hooks, 2000; LaFrance & Wigginton, 2019; Tandon, 2018).

A critical intersectional feminist approach is not without its own challenges. For instance, intersectionality is often critiqued for presenting marginalised groups such as the LGBTQIA+ community as a uniform identity with a single and shared lived experience (Gringeri, Wahab, & Anderson-Nathe, 2010). To counter this, it is important not to essentialise experience and identity but acknowledge "the complexities of multiple, competing, fluid, and intersecting identities" (ibid., p.394). Therefore, designing technology in a participatory manner with people with multiple lived experiences, including vulnerable groups, is so essential to the development of digital solutions which seek to do true 'social good' without profiting off a crisis.

An additional recommendation is to employ critical reflexivity into one's work, to enable individuals to reflect on their positionality, power, and privilege when developing and designing technology. This helps to interrogate not only one's role in the development and design but also the ways in which power is distributed and plays out (Cooky et al., 2018). Reflexivity acknowledges that people are not separate from their work and that they impose or bring their values to their work, whether they are researchers or software developers. Those developing digital solutions to a crisis are not separate from the design or development process, but their values, politics, personal identities, and assumptions about the world influence and underpin the technology they develop.

In their 2019 report on feminist internet ethical research practices, the APC proposes that a feminist ethics of care should extend to digital security and safety. This includes accounting for digital vulnerabilities such as security breaches or the leaking of personal data. An ethics of care is informed by feminist values and emphasises "care and responsibility rather than outcomes" (Edwards & Mauthner, 2012, p.19). Additional areas of consideration include representation, language, experience, and subjectivity which makes such an approach well-suited to considering the needs of marginalised groups (Blakely, 2007, pp.64-5). In adopting a feminist ethics of care approach to the deployment of digital solutions, the well-being of potential adopters, including LGBTQIA+ people, is at the forefront of the development process.

## 6.2. Decentralised technology and user-controlled access

A critical intersectional feminist approach disrupts oppressive power structures and favours the decentralization of technology. In this manner, the technology is "under control of independent responsible authorities, would evolve and be open source" (Desai, 2020). Independent responsible authorities would be community-appointed and will include those who are often left out of decision-making processes. Access to a user's data should only be allowed if the owner of the data "approves it, for each use" (ibid.). Continuing, Desai (2020) writes, "If stored, for backup it should not be disclosed to third parties and should be deleted when requested. There must be a legal requirement for the length".

It is not enough for users to opt in voluntarily, but they must also be made fully aware of the implications for handing over access to their data. The length of terms of service documents, wordiness, and the use of jargon often result in users agreeing or consenting to terms "without fully understanding how their safety - their lives - may be at risk" (Nigro, 2019). Consequently, terms of service or terms and conditions need to be made far more accessible to the average person.

Chair (2020) writes on how some national data protection authorities have done well to provide guidance on data protection. These authorities "provided pointers to avoid future misuse of technology such as unwarranted increased surveillance and determine who has authority in collection and processing of data" (ibid.). In Senegal and in Mauritius, for example, data protection authorities ensured and assessed the anonymization of data for "statistical and scientific research purposes to ensure the data may not be de-anonymized" (ibid.), or that those who granted access to their data were duly informed and a time limitation on access to this data was in place.

Ultimately, digital contact tracing, driven by Big Tech as "self-proclaimed" health policymakers, leaves untenable room for exploitation and harm. This is especially the case for LGBTQIA+ persons, who may be in physical danger should their whereabouts, identities, or personal views be exposed. The widespread trade-off between individual privacy (or freedom) and public health is unjust in the broader context of marginalization, and further disenfranchises those at the peripheries of society. A critical intersectional feminist approach can mitigate some of these concerns by proposing solutions that are more equal, inclusive, and socially just. Such an approach, informed by a feminist ethics of care, is transparent, user-centric, and sensitive to volatile political and cultural dynamics. We encourage future research that seeks to operationalise a critical intersectional feminism in the context of public health.

# REFERENCES AND CITATIONS